\documentclass[a4paper]{jpconf}
\usepackage{graphicx}
\usepackage{hyperref}
\usepackage{cite}
\usepackage{lineno}

\begin{document}
\title{The Application of SNiPER to the JUNO Simulation}

\newcommand{\IHEP}{$^1$}
\newcommand{\SCU}{$^2$}
\newcommand{\SDU}{$^3$}
\newcommand{\SYSU}{$^4$}

\author{Tao Lin\IHEP, Jiaheng Zou\IHEP, Weidong Li\IHEP, Ziyan Deng\IHEP, Xiao Fang\SCU, Guofu Cao\IHEP, Xingtao Huang\SDU{} and Zhengyun You\SYSU{}}
\author{(On Behalf of the JUNO Collaboration)}

\address{\IHEP Institute of High Energy Physics, Chinese Academy of Sciences, Beijing, China}
\address{\SCU Sichuan University, Chengdu, China}
\address{\SDU Shandong University, Jinan, China}
\address{\SYSU Sun Yat-sen University, Guangzhou, China}

\ead{lintao@ihep.ac.cn}

\modulolinenumbers[1]

\begin{abstract}
The JUNO (Jiangmen Underground Neutrino Observatory) is a multipurpose neutrino experiment which is designed to determine neutrino mass hierarchy and precisely measure oscillation parameters. As one of the important systems, the JUNO offline software is being developed using the SNiPER software. In this proceeding, we focus on the requirements of JUNO simulation and present the working solution based on the SNiPER.

The JUNO simulation framework is in charge of managing event data, detector geometries and materials, physics processes, simulation truth information etc. It glues physics generator, detector simulation and electronics simulation modules together to achieve a full simulation chain. In the implementation of the framework, many attractive characteristics of the SNiPER have been used, such as dynamic loading, flexible flow control, multiple event management and Python binding. Furthermore, additional efforts have been made to make both detector and electronics simulation flexible enough to accommodate and optimize different detector designs.

For the Geant4-based detector simulation, each sub-detector component is implemented as a SNiPER tool which is a dynamically loadable and configurable plugin. So it is possible to select the detector configuration at runtime. The framework provides the event loop to drive the detector simulation and interacts with the Geant4 which is implemented as a passive service. All levels of user actions are wrapped into different customizable tools, so that user functions can be easily extended by just adding new tools. The electronics simulation has been implemented by following an event driven scheme. The SNiPER task component is used to simulate data processing steps in the electronics modules. The electronics and trigger are synchronized by triggered events containing possible physics signals.

The JUNO simulation software has been released and is being used by the JUNO collaboration to do detector design optimization, event reconstruction algorithm development and physics sensitivity studies.

\end{abstract}

\section{Introduction}

JUNO \cite{An:2015jdp,Djurcic:2015vqa} is a multi-purpose neutrino experiment designed to measure neutrino mass hierarchy and precisely measure oscillation parameters and detect astrophysical and geological neutrinos. It will be located in southern China about 53~km away from Yangjiang and Taishan nuclear power plants.

\begin{figure}[h]
\begin{center}
\includegraphics[width=20pc]{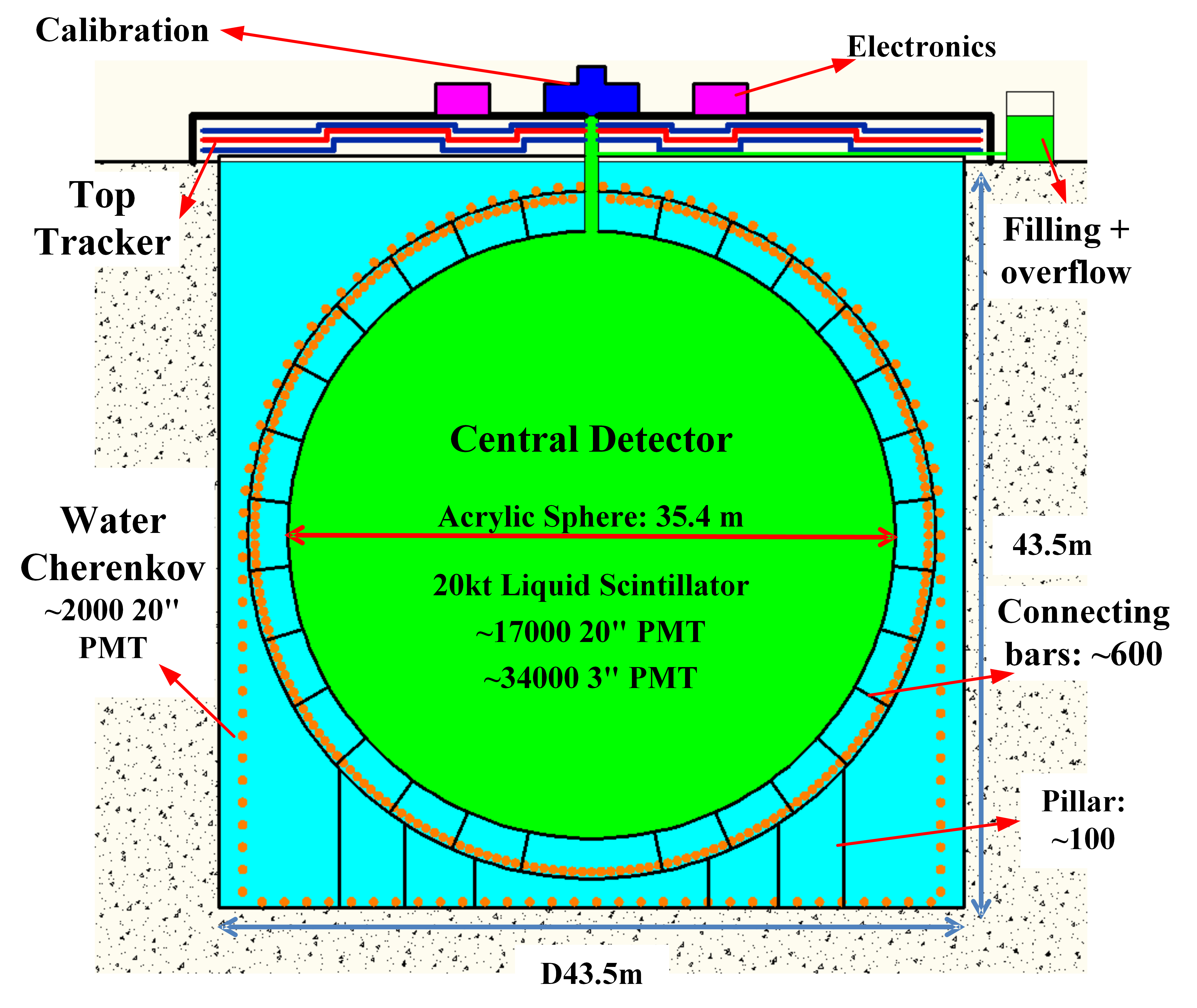}\hspace{2pc}%
\caption{\label{fig:10}Schematic view of JUNO detector}
\end{center}
\end{figure}

Figure~\ref{fig:10} shows the schematic view of the JUNO detector. The central detector (CD), which is the innermost part, is used for neutrino detection. It consists of a spherical acrylic vessel containing 20~kt liquid scintillator (LS). Photomultiplier tubes (PMTs) are placed surrounding the acrylic vessel to collect light from LS. Around the CD, a water pool is used to shield radioactivities. PMTs in the water pool are used to veto cosmic ray muon events by detecting Cherenkov light. On the top of the water pool, there is a top tracker which is also used to measure and veto muons.

\begin{figure}[h]
\begin{center}
\includegraphics[width=20pc]{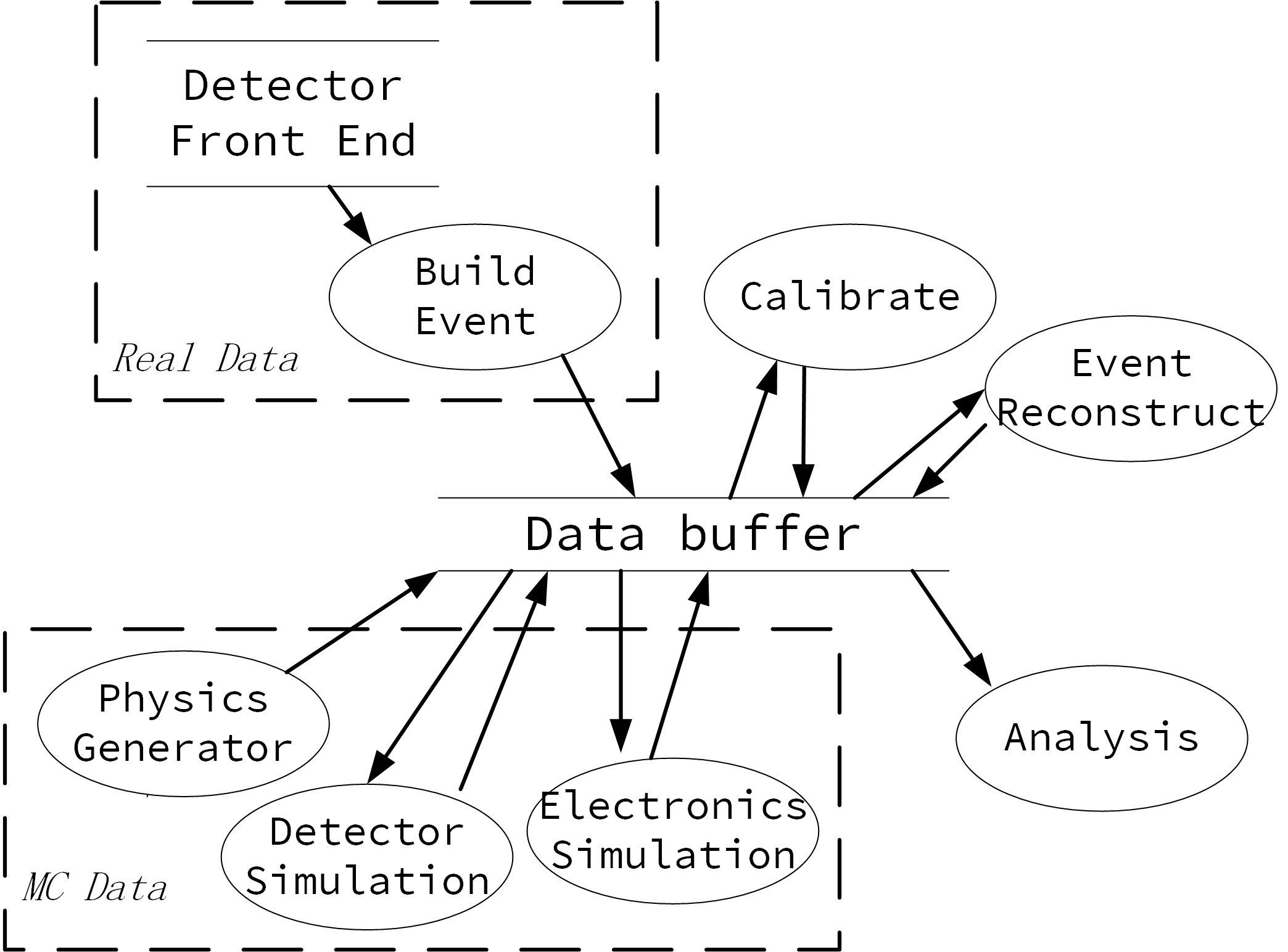}\hspace{2pc}%
\caption{\label{fig:11}Work flow of data processing}
\end{center}
\end{figure}

The raw data needs to be processed offline and reconstructed for further physics analysis. The simulation data and raw data share the same event data model, so the simulation data could be processed in the same way. Figure~\ref{fig:11} shows the current data processing steps, which are implemented as different algorithms following the scheme of the SNiPER \cite{Zou:2015ioy} framework. Each algorithm reads event data \cite{Li:edm} from data buffer and does corresponding calculations, then puts the result back to data buffer. A ROOT I/O \cite{Brun:1997pa} service is used for event data persistency from data buffer to disk and vice versa. A detector geometry service uses GDML \cite{Chytracek:2006be} and ROOT as detector geometry description, which can be used by algorithms in a unified way. 

In Figure~\ref{fig:11}, the physics generators generate kinematic information of primary particles, which are saved into GenEvent objects. The format of HepMC \cite{Dobbs:2001ck} is used in GenEvent. Other formats, such as HepEvt and GENIE \cite{Andreopoulos:2009rq}, are converted to HepMC via corresponding tools. In the next step, the detector simulation algorithm accesses these GenEvent objects and starts tracking. Hits, which contain charge and time information, are generated in sensitive detectors and saved in SimEvent objects. After that, the electronics simulation algorithm reads these SimEvent objects and performs the digitization, which generates ElecEvent objects containing waveforms information. These waveforms are processed by PMT calibration algorithm and CalibEvent objects are saved. The event reconstruction algorithm performs the event reconstruction by reading CalibEvent objects and stores RecEvent objects. At last, physicists can perform any physics analysis from RecEvent objects.

As part of the offline data processing software \cite{Huang:junooffline}, the reliable Monte Carlo (MC) simulation software plays an important role for detector parameters optimization and physics study. A simulation framework is developed based on SNiPER framework, consisting of physics generators, detector simulation and electronics simulation modules. It is integrated with SNiPER and takes advantages of SNiPER. 

\section{Detector simulation}

The JUNO detector simulation software, which is based on Geant4 \cite{Agostinelli:2002hh,Allison:2006ve}, was developed as a standalone application. In order to run simulation in the framework, several interface classes are designed to integrate SNiPER and Geant4 without too much modification to the user code. The extendibility supplied by SNiPER simplifies management of physics generator, detector geometry construction, user actions and so on. At the same time, tools are used to configure parameters using Python scripts instead of using Geant4's macro files.

\begin{figure}[h]
\begin{center}
\includegraphics[width=25pc]{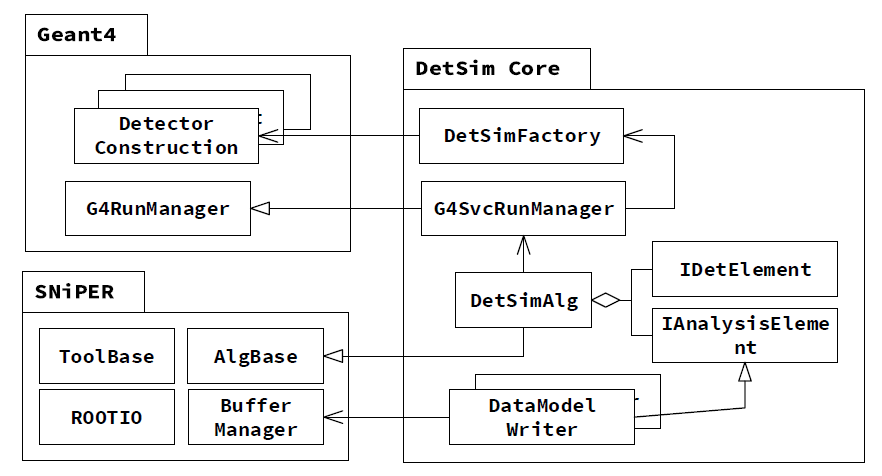}\hspace{2pc}%
\caption{\label{fig:20}Class diagram of detector simulation.}
\end{center}
\end{figure}

Figure~\ref{fig:20} shows the design of detector simulation framework. This detector simulation framework is a bridge between Geant4 and SNiPER. It consists of the integration of Geant4 and SNiPER, a configurable user interface, geometry management and modularized user actions. With this extra layer as middleware, detector simulation becomes a seamless component of the offline software. 

Both SNiPER and Geant4 have their own event loop control. It's necessary to adapt Geant4's event processing work flow into SNiPER. In order to adopt SNiPER's event loop, an algorithm named {\tt DetSimAlg} is invoked event by event in SNiPER. Then {\tt DetSimAlg} invokes a service named {\tt G4SvcRunManager}, which is derived from {\tt G4RunManager}, to initialize simulation and start tracking.
To decouple run manager and user code, such as detector construction and physics list, a factory class named {\tt DetSimFactory} is implemented. This class is responsible for constructing user code and passing them to {\tt G4SvcRunManager}.

Physics generator interfaces, physics processes and detector components are all configurable. Instead of using Geant4's macro files, they are configured by lightweight tools. This keeps user interface same as other algorithms by using Python script only. To be compatible with Geant4's macro files, a command line option can be used to specify input macro files. For example, user can visualize detectors by specifying a visualization macro file.


Geometry management is an issue when there are several different detector design options. To be consistent and flexible, a detector element is proposed to represent one detector component, which is a high level concept. For example, a central detector is a detector element, which contains LS, container and buffer material only. However, PMTs are not placed in this detector element. Therefore different options of central detector can be interchanged while keeping the arrangement of PMTs unchanged. It's also useful when different types of PMTs are used. These PMTs are managed in a unified way. Besides real detector components, a detector component based on GDML can construct any geometries and materials via parsing an input GDML file. There is another interface to control the placement of detector elements. Calibration units are placed into the central detector at runtime by a corresponding configuration. 


User actions are important for developers to get MC truth information during detector simulation. To modularize these user actions, they are organized into a list of lightweight tools, which can access all necessary information supplied by Geant4. A manager, which manages these tools, is invoked by different levels of user actions. Then it dispatches calculation to these tools according to their registration order. Users can load the corresponding tools according to requirements. Developers can create a new tool without modifying any other tools. There are some ``official'' tools loaded by default, such as writers for event data model and geometry.


\section{Electronics simulation} 

An IBD (inverse beta decay, $\bar{\nu}_e + p \to e^+ + n$) event consists of a prompt signal from $e^+$ and a delayed neutron captured on hydrogen. Electronics simulation needs to split such a physics event into two readouts. Meanwhile, detector backgroud mixing is also needed. In order to handle both event splitting and event mixing, a ``pull'' mode is proposed. 

\begin{figure}[h]
\begin{center}
\includegraphics[width=35pc]{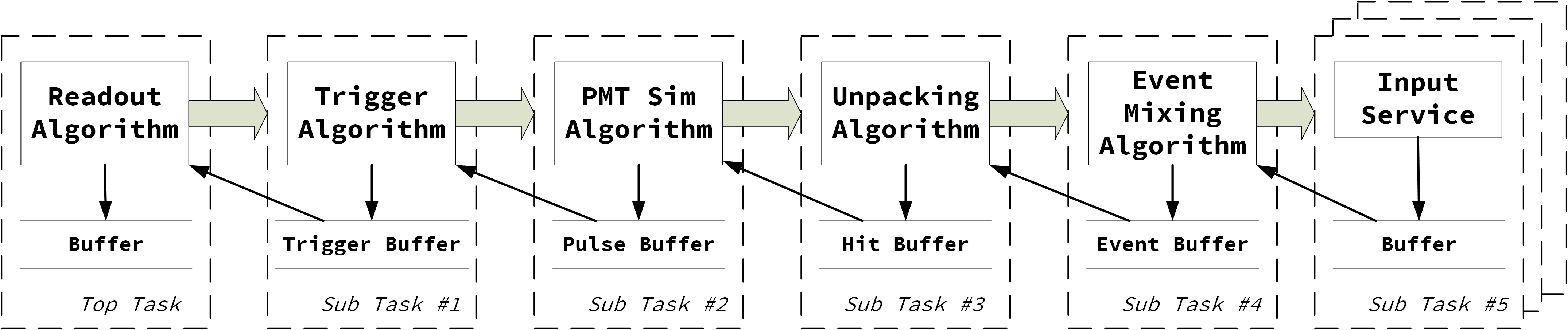}\hspace{2pc}%
\end{center}
\caption{\label{fig:30}Work flow of electronics simulation}
\end{figure}

Instead of starting from an event mixing algorithm, electronics simulation starts from a readout algorithm, as shown in Figure~\ref{fig:30}. At first, the readout algorithm needs a trigger from the trigger buffer to create a readout event. If the corresponding trigger buffer is empty, the trigger simulation algorithm is asked to execute. Then this trigger simulation needs enough pulses to give a trigger of an event. If there are not enough pulses in the pulse buffer, a PMT simulation algorithm starts to produce pulses from the hit buffer, until there are enough pulses in a time window. An unpacking algorithm unpacks hits from different events into the hit buffer, while an event mixing algorithm put events into the event buffer. After all buffers are filled with corresponding data, the readout algorithm digitizes the waveforms and saves them into readout event. In such a process, simulation of one event is done. Finally the global time stamp is updated to synchronize the data buffers. For memory optimization, circular buffers are used in some algorithms, such as the waveform algorithm.

In this implementation, each algorithm is associated with a task. These tasks are executed passively and invoked only when the data in the corresponding buffer is not enough. SNiPER's incident mechanism is used to communicate between an algorithm and a task. When a task is invoked, registered algorithms in this task are executed once. At the same time, input service, output service and buffer service of this task are updated. To mix events from different files, tasks with only input service and buffer service are setup dynamically. User can also use a label name to identify type of data and associate its event rate. According to the event rate, an event mixing algorithm invokes corresponding tasks respectively. Then this event mixing algorithm accesses buffers in these tasks and puts events into its own buffer. Then an unpacking algorithm sorts all hits by time order and mixes them up. In this way, hit-level event mixing is achieved.

\section{Performance measurements}

IBD events are used for performance study. Detector simulation, electronics simulation without mixing and electronics simulation with mixing are measured respectively. For each category, 20 jobs with 1000 readout events per job are prepared. Jobs were submitted to a dedicated blade server with the Intel Xeon E5-4620 v2 @ 2.60GHz CPU and 132 GB of memory. When a job is running, a daemon process monitors its CPU and memory usage and saves the result into files. To eliminate interference, this daemon updates the result every 5 seconds.

Due to readout splitting of IBD events, one SimEvent object will produce two ElecEvent objects. For an electronics simulation job, 1000 readout objects are produced from less than 500 IBD events.

Hit-level event mixing is performed in electronics simulation only. The input SimEvent objects are prepared before event mixing. For the test, 1~Hz of IBD events are used here, with 6~Hz of U-238 events, 6~Hz of Th-232 events and 2~Hz of K-40 events. The other configuration is the same as previous measurements.


\begin{table}[h]
\caption{\label{tab:perf}Summary of software performance}
\begin{center}
\begin{tabular}{lrrr}
\br
 & DetSim & ElecSim & ElecSim \\
  &  & w/o mixing& w/ mixing\\
\mr
Physical memory (GB) &   0.9 &  0.5 &  0.6 \\
Virtual memory (GB)  &   1.4 &  2.2 &  2.4 \\
CPU time (min)       & 168.9 & 21.2 & 14.2 \\
CPU utilization (\%) &  98.9 & 94.2 & 92.2 \\
Size per event (MB)  &   0.1 &  5.8 &  3.5 \\
\br
\end{tabular}
\end{center}
\end{table}

Table~\ref{tab:perf} shows the summary of simulation software performance. Detector simulation is both time and memory consuming, due to propagation of thousands of optical photons. Its CPU utilization shows that detector simulation software is efficient even though integrated with SNiPER. For high energy cosmic ray muon events, millions of optical photons are generated in liquid scintillator, some methods are studied to speed up the simulation \cite{Lin:2016vua}. 

Electronics simulation is memory and I/O consuming. Even though with optimization of data structure, thousands of waveforms with 1 GHz of sampling rate still occupy a lot of memory and disk usage. For event mixing, it takes less time and disk usage due to lower energies of particles from background. Most readout objects are actually background only. However, event mixing needs to load events from multiple files randomly, so CPU utilization is suppressed due to I/O latency.

\section{Conclusions}

In this proceeding, we show how SNiPER is applied in the JUNO simulation software, including physics generator, detector simulation and electronics simulation modules. By introducing several middleware classes, SNiPER and Geant4 are integrated to simplify the management of geometry and user actions. Via incident mechanism in SNiPER, electronics simulation is implemented in a ``pull'' work flow, which supports hit level mixing naturally. The performance of current simulation software shows the software fulfills requirements as well.
 
However, there are still several challenges, such as simulation of huge number of optical photons and PMTs. The concurrent computing using GPU and phi-coprocessor is being studied in order to speed up the simulation of light propagation in the large liquid scintillator detector.

\section*{Acknowledgements}
This work is supported by Joint Large-Scale Scientific Facility Funds of the NSFC and CAS (U1532258), the Strategic Priority Research Program of the Chinese Academy of Sciences, Grant No. XDA10010900, National Natural Science Foudation of China (11575224, 11405279, 11675275).

\section*{References}
\bibliography{iopart-num}

\end{document}